\documentclass[aps,prl,twocolumn,showpacs,superscriptaddress,groupedaddress,
nofootinbib]{revtex4}  
\usepackage{graphicx}  
\usepackage{dcolumn}   
\usepackage{bm,amsmath}        
\usepackage{amssymb}   
\usepackage{color}

\newcommand{\beq}{\begin{equation}} 
\newcommand{\eeq}{\end{equation}} 
\newcommand{\bea}{\begin{eqnarray}} 
\newcommand{\eea}{\end{eqnarray}}

\begin{document}

\title{Discoveries far from the Lamppost with Matrix Elements and Ranking}
\author{Dipsikha Debnath} \affiliation{Physics Department, University
  of Florida, Gainesville, FL 32611, USA}
\author{James~S.~Gainer} \affiliation{Physics Department, University
  of Florida, Gainesville, FL 32611, USA}
\author{Konstantin~T.~Matchev} \affiliation{Physics Department,
  University of Florida, Gainesville, FL 32611, USA}
\date{May 22, 2014}

\begin{abstract}
The prevalence of null results in searches for new physics at the LHC
motivates the effort to make these searches as model-independent as
possible.  We describe procedures for adapting the Matrix Element
Method for situations where the signal hypothesis is not known \emph{a
  priori}.  We also present general and intuitive 
approaches for performing analyses and presenting results, which
involve the flattening of background distributions using likelihood
information.
The first flattening method involves ranking events by background
matrix element, the second involves quantile binning with respect to
likelihood (and other) variables, and the third method involves
reweighting histograms by the inverse of the background distribution.
\end{abstract}

\maketitle
{\bf Introduction.}~~ 
The CERN Large Hadron Collider (LHC) will soon resume operation,
probing energies never before accessed with colliders.  Ideas about
what sort of new physics it will discover often focus on models that
resolve the hierarchy problem~\cite{hierarchy} or provide relic dark
matter candidates~\cite{DM}.  There are a great variety of ideas in
each category.
However, the lack of convincing evidence for 
new physics at the LHC to date suggests that we may
be looking in the wrong places.  We therefore consider methods
that will allow the discovery of \emph{any} departure from known
physics.

The Matrix Element Method (MEM)~\cite{MEM} and similar 
multivariate analyses~\cite{MVA}
have been used with success in LHC experiments.  A particularly
dramatic example has been the use of the MEM in the four-lepton
channel for the discovery of the Higgs Boson~\cite{Higgs Discovery}
and the measurement of its properties~\cite{Higgs Properties}.  Such
analyses used variables, such as MELA KD~\cite{MELA} or
MEKD~\cite{MEKD} that involve the ratio of signal and background
matrix elements.  Clearly these variables are, therefore, optimized to
the appropriate signal and background hypotheses.  

A natural question is whether we can preserve, to some degree, the
sensitivity of the MEM without assuming a specific signal hypothesis.
We argue that this is possible.  
Specifically, one can use the log likelihood of events, calculated
using the likelihood for background events, as a test statistic in a
way that allows a model-independent discovery of new physics.  We then
demonstrate how the optimal MEM-based (as well as other) techniques can be used to create
flat distributions for the background with respect to kinematic
variables of interest, including, most importantly, variables derived
from the matrix element.  

{\bf The Matrix Element Method.}~~
According to the Neyman-Pearson lemma~\cite{NP}, the optimal test
statistic for comparing hypotheses $H_0$ and $H_1$ is provided by the likelihood ratio:
\begin{equation}
\label{eq:ratio}
R_\Lambda(H_0,H_1) = \frac{\Lambda_{H_0}(\{ {\mathcal E}_i\})}{\Lambda_{H_1}(\{{\mathcal E}_i\})},
\end{equation}
where $\Lambda_{H_0} (\Lambda_{H_1)}$ is the likelihood for the hypothesis $H_0(H_1)$ as
a function of the data, which we assume consists of $N$ events, ${\mathcal E}_i$.  
In the MEM, the likelihood for a given event is calculated using the
expression
\begin{equation} \label{eq:mem1}
{\cal P}({\mathcal E}_i |H_i) = \frac{1}{\sigma(H_i)}
\sum_{k,l} \int dx_1 dx_2 \, \frac{f_k(x_1)f_l(x_2)}{2sx_1x_2} \\
\end{equation}
\vspace{-16 pt}
\begin{equation*}
\times \biggl [ \prod_{\text{~all~}j} \int
 \frac{d^3 q_j}{(2\pi)^3 2E_j} \biggr] 
\times \biggl[ \prod_{\text{~visible~}j}  T(\{q_j\},\{p_j\}) \biggr]
\end{equation*}
\vspace{-16 pt}
\begin{equation*}
 \times
|{\cal M}_{H_i, kl}(\{q_j\})|^2,
\end{equation*}
where ${\mathcal M}_{H_i, kl}$ is the theoretical matrix element for hypothesis
$H_i$, $f_k$ and $f_l$ are parton distribution functions (pdf) as a
function of momentum fractions $x_1$ and $x_2$, while $\sigma(H_i)$ is
the total cross section after acceptances, efficiencies, etc.  The
likelihood for a set of $N$ events $\{\{p_i^{\rm vis}\}_j\}$ (where $j$
ranges from $1$ to $N$) is simply the product of the likelihoods for
each event:
\begin{equation}
\Lambda_{H_i}(\{ {\mathcal E}_i \}) = \prod_i^N {\cal P}({\mathcal E_i}|H_i).
\end{equation}
Thus the likelihood ratio (\ref{eq:ratio}) contains the product of ratios of
event-by-event likelihoods described in Eq.~(\ref{eq:mem1}).  Often,
the two hypotheses ($H_0$ and $H_1$) will involve the same final state,
hence factors due to the phase space integrals in Eq.~(\ref{eq:mem1})
will cancel in the likelihood ratio.  We are then left with a ratio of squared
matrix elements, possibly weighted (in the case where the hypotheses
involve different initial state partons) by pdfs.  These squared
matrix elements contain a great deal of information about the process,
including the pole structure, spin correlations, etc.  While the
implementation of an analysis using the likelihood ratio (\ref{eq:ratio}) as a
test statistic may sometimes be challenging in practice, conceptually the
implementation is straightforward and the sensitivity is, at least in
principle, optimal \cite{MELA,MEKD}. 

{\bf Discovery from Background Likelihood Distributions.}~~
The limitation of the MEM is that we must know the signal process in 
order to calculate the appropriate likelihood.
As a result, if we do not know what signal model 
we are looking for we can no longer consider the likelihood
ratio, as we know only one hypothesis, the background. 
It will still be useful,
however, to use the information about
the background that is encoded in the matrix element.  Therefore, we
propose that we consider the background likelihood,
\begin{equation}
\label{eq:test-bg}
\Lambda_B = \prod_{i=1}^N P({\mathcal E}_i|~{\rm bg}),
\end{equation}
and closely-related expressions, as test statistics.
Here $N$ is the number of events and $P({\mathcal E}_i|~{\rm bg})$ is
either defined following Eq.~(\ref{eq:mem1}) for the background
hypothesis or is a similar variable.  
Thus, in this letter, our test statistic will be the sum of the
logarithm of the pdf-weighted squared background matrix elements, as a
function of the visible momenta in each event in our data sample,
calculated using~{\sc MEKD}~\cite{MEKD} (a package for MEM
calculations for the four-lepton final state based on {\sc
  MadGraph}~\cite{MadGraph}). 
The event-by-event value  of this quantity will be labelled $|{\mathcal M}|^2$ in subsequent figures, 
while the sum of this quantity over events in the pseudo-experiment
will be labelled $\Lambda_B$.

\begin{figure}[t]
\includegraphics[width=\columnwidth]{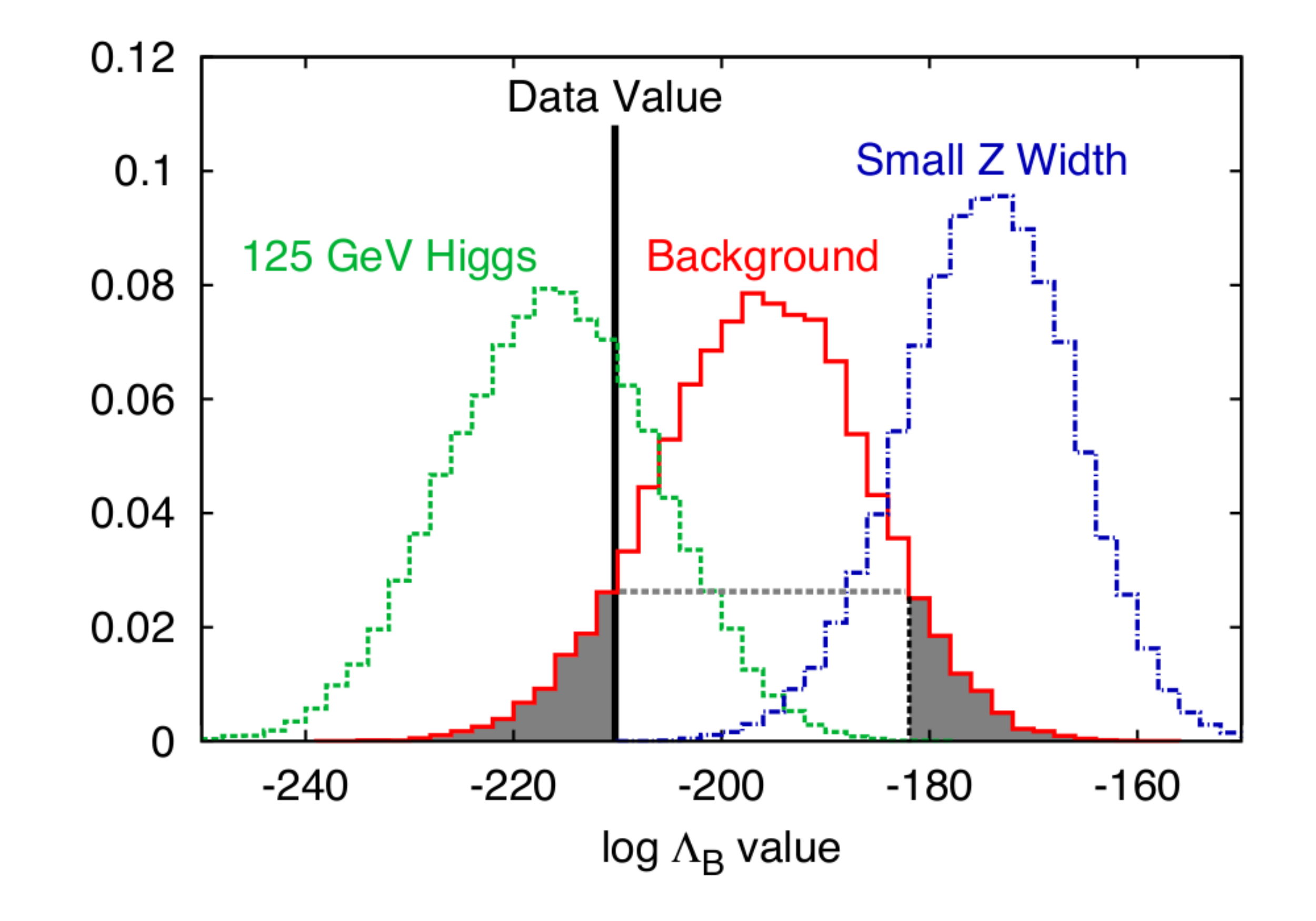} \\
\vspace{-15 pt}
\caption{\label{fig:log-mekd-pseudos}
The unit normalized distribution of our test statistic, defined in
Eq.~(\ref{eq:test-bg}) as evaluated for 20-event pseudoexperiments
consisting of background $q\bar{q} \to 4\ell$ events (red solid
curve), $gg \to H \to 4\ell$ signal events for a $125$ GeV Higgs
(green dashed curve), and $q\bar{q} \to 4\ell$ events for which the
$Z$ boson width has been reduced by a factor of $5$ (blue dot-dashed
curve).  If a particular value of our test statistic, indicated by
``Data Value'', is observed, the corresponding $p$-value is given by
the area in gray.  In this specific case, $p \approx 0.13$.
}
\end{figure}

As an example, we plot the distribution of $\Lambda_B$ in
Fig.~\ref{fig:log-mekd-pseudos}, where we show this quantity as
calculated for psuedo-experiments consisting of $20$ events, generated
using the indicated hypotheses, namely the irreducible $q\bar{q} \to
2e2\mu$ background (red solid curve), gluon fusion production of a
$125$ GeV Higgs boson that decays to $2e2\mu$ (green dashed curve),
and the irreducible  $q\bar{q} \to 2e2\mu$ with the $Z$ boson width
scaled down by a factor of $5$ (blue dot-dashed curve).
 
In this figure, we also demonstrate the procedure for obtaining a
$p$-value, describing the extent to which actual data is consistent
with the background hypothesis.  
(This is a somewhat ``brute force'', but conceptually straightforward
approach to the problem of evaluating the goodness of fit of 
a likelihood~\cite{Likelihood Fit}.)
Specifically, one takes the particular value of $\Lambda_B$
measured in the data, labeled ``Data Value'', following
Eqs.~(\ref{eq:mem1}) and~(\ref{eq:test-bg}) and evaluates the fraction
of background pseudo-experiments which have values of $\Lambda_B$ which
are of equal or lesser likelihood than ``Data Value''.   
This corresponds to the shaded region in the figure.
The specific value of ``Data Value'' used in this figure is
consistent with the hypothesis of a $125$ GeV Higgs boson, though many
other signal hypotheses produce events with lower values of $\Lambda_B$ than
would be expected from background events.  However, it is also
possible to have a signal hypothesis that is ``more background-looking" than the
background itself, such as the hypothesis of background events with a
reduced $Z$ boson width, as can be seen from the blue dot-dashed
curve.

{\bf How to Flatten Background Distributions: Examples.}~~
The main point of this letter is that the procedure above allows one
to exclude the background hypothesis in the presence of an {\em unknown}
signal.  In other words, one can confidently look for new physics models
``away from the lamppost", i.e., models which no theorist has yet thought of. 
While, in principle, any variable could have been used to
construct such a test statistic, the use of a variable based on the
background likelihood should additionally optimize the sensitivity of such searches.

We now present some related methods, which allow the 
``non-backgroundness'' of some potential signal to be shown in a 
clear and intuitive way.  These methods also have the benefit that they
generalize to any possible channel, so results and sensitivity 
in various channels can easily be compared.

\begin{figure}[t]
\includegraphics[width=0.98\columnwidth]{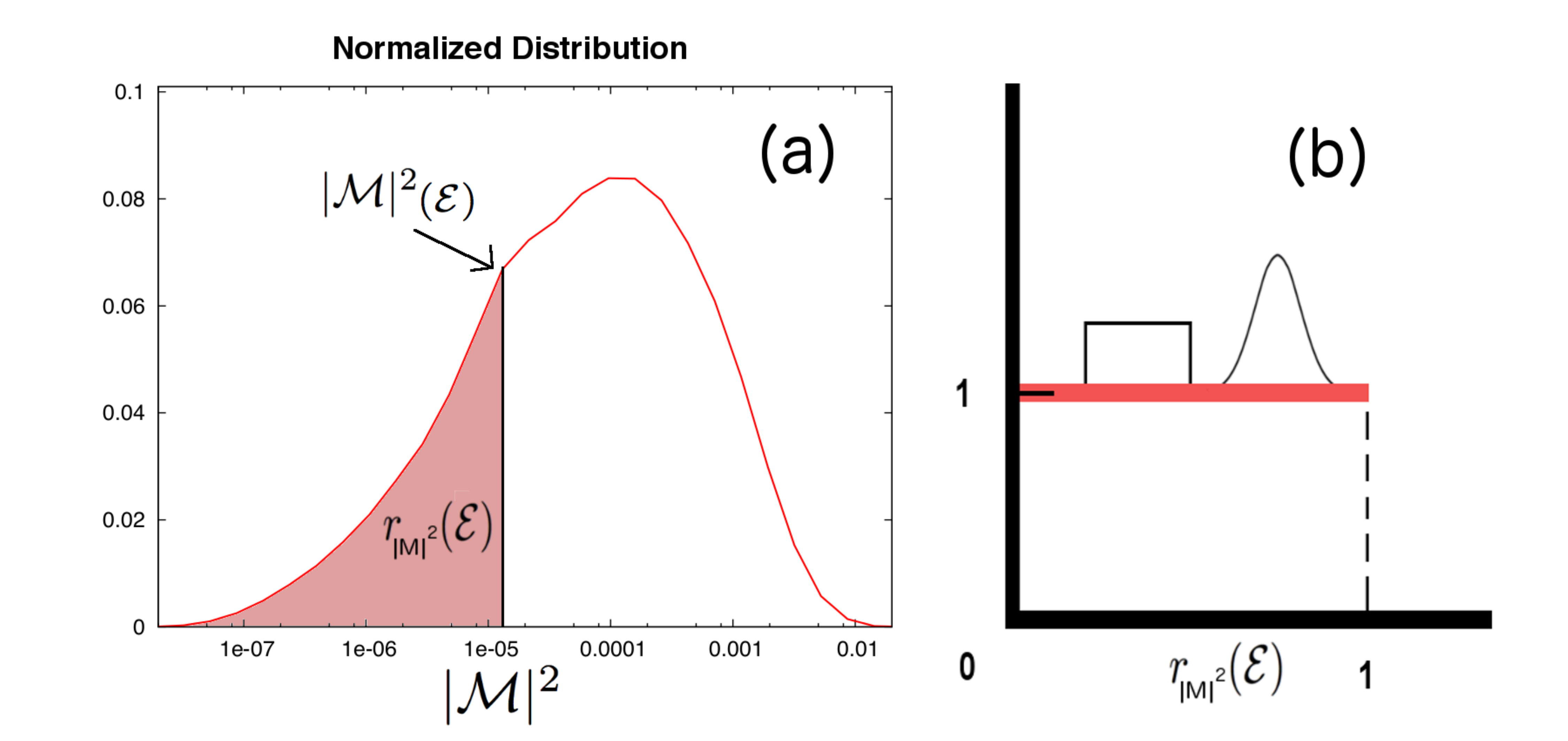}\\
\caption{\label{fig:ranking-cartoons}
Panel (a) shows how the distribution of background MEKD for background
events is used to create a ``ranking variable".
In panel (b) we show that the background distribution with respect to
this ranking variable is flat, while for other processes the
distribution of background ranking variable is not flat.}
\end{figure}

\emph{1.~~Flattening with Ranking}.~~
In this approach, one takes the normalized distribution,
$\frac{dN}{d\xi}$, for some kinematic variable, $\xi$,
and defines a ``ranking'' variable,
\begin{equation}\label{eq: rank int}
r(\xi)=\int_{-\infty}^\xi \frac{dN}{d\xi^\prime} d\xi^\prime.
\end{equation}
We note that $r(\xi)$ is the cumulative distribution function for
the background with respect to the variable $\xi$.  
We can now evaluate the ranking $r_\xi$ of any given event ${\mathcal E}$
by defining
\begin{equation}
r_\xi ( {\mathcal E}) = r (\xi( {\mathcal E})),
\end{equation}
that is, the value of the ranking variable for a given event, 
${\mathcal E}$, is the value found from Eq.~(\ref{eq: rank int})
for the value of the kinematic variable $\xi$ obtained for the 
event.  The connection between our ranking variable $r_\xi ( {\mathcal E})$ 
and the background $\xi$ distribution is shown pictorially in
Fig.~\ref{fig:ranking-cartoons}(a). The figure also illustrates 
the physical meaning of $r_\xi ( {\mathcal E})$ --- it is the 
fraction of background events ${\mathcal E}'$ in which 
$\xi({\mathcal E}')<\xi({\mathcal E})$. 
If we then consider the normalized distribution
of the background with respect to $r_\xi$, we find that 
\begin{equation}
\frac{dN}{dr_\xi} = 1,
\end{equation}
hence the distribution of this variable for background events is flat,
as is shown in Fig.~\ref{fig:ranking-cartoons}(b).
This procedure, of course, works for any kinematic variable, $\xi$, 
we especially recommend using it with the sensitive 
matrix-element-based variables advocated above.
Thus, one obtains a sensitive variable for which the background
distribution is flat, while the distribution of signal events is 
characterized by departures from flatness, as is also shown in
Fig.~\ref{fig:ranking-cartoons}(b).


We note in passing that calculating $r(\xi)$ from Monte Carlo (MC)
events is quite straightforward.  One simply calculates the value 
of the variable $\xi$ for each of the $N$ events in the MC sample, 
thus obtaining a list of values $\{\xi_i\}$.  
The value of $r(\xi)$ is then well-approximated by the fraction 
of the $\{\xi_i\}$ which are less than $\xi$, i.e.,
\begin{equation}
  r(\xi) \approx \frac{1}{N} \sum_{i=1}^N \theta(\xi-\xi_i).
\end{equation}
This procedure should facilitate the experimental implementation of
this technique.

\begin{figure}[t]
\includegraphics[height=0.435 \columnwidth]{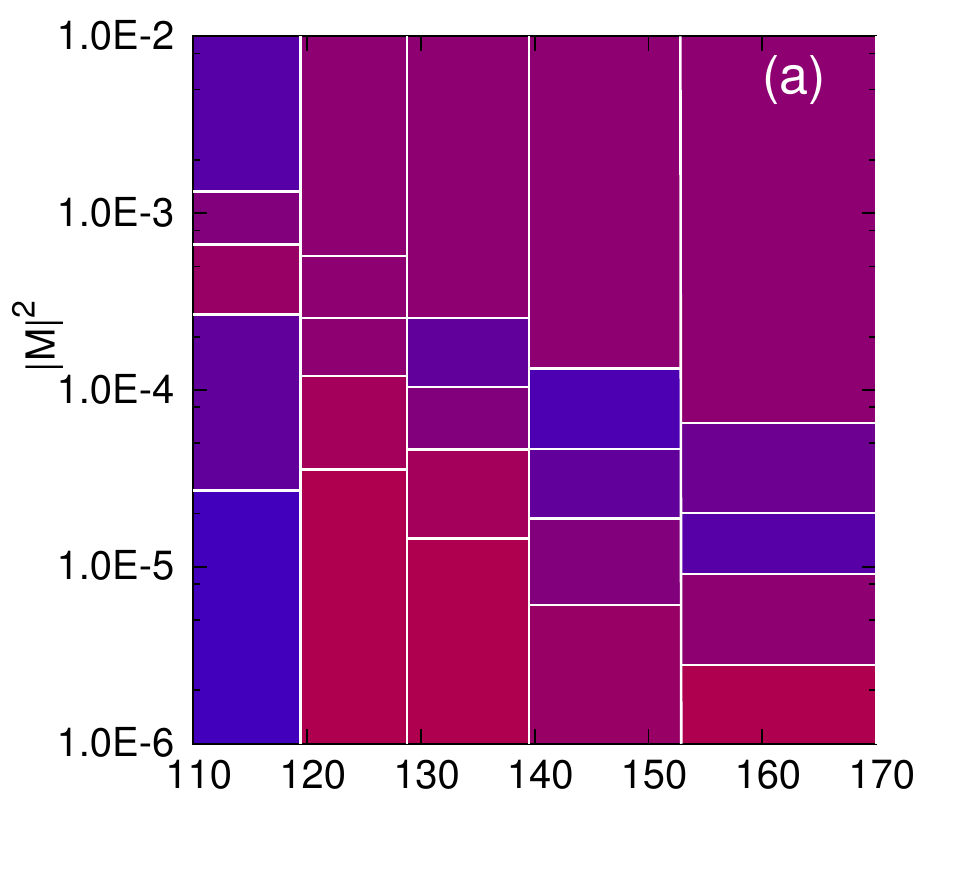} 
\includegraphics[height=0.435 \columnwidth]{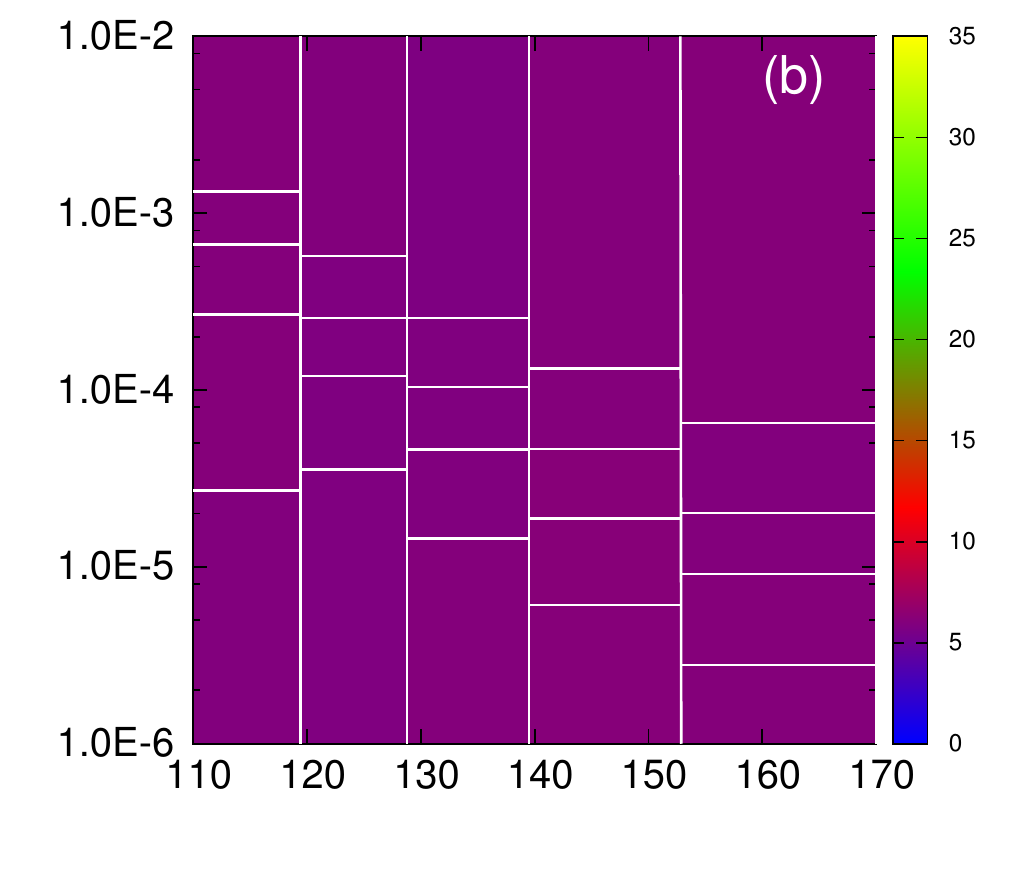}\\
\includegraphics[height=0.435 \columnwidth]{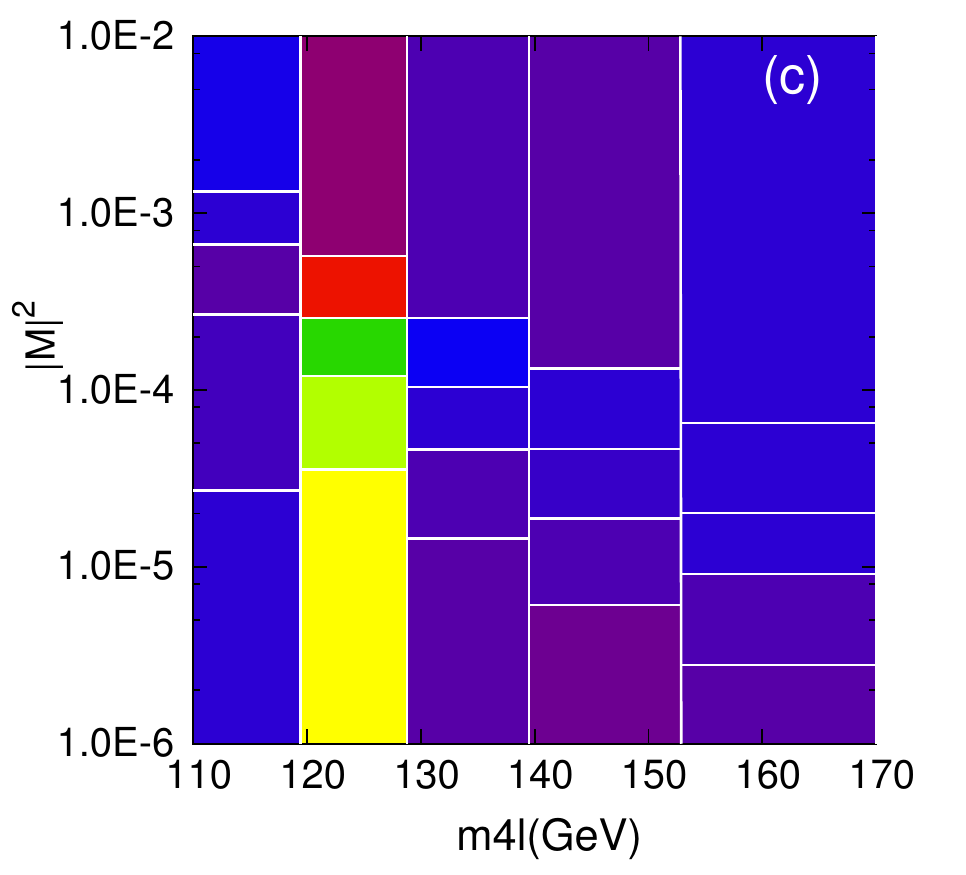} 
\includegraphics[height=0.435 \columnwidth]{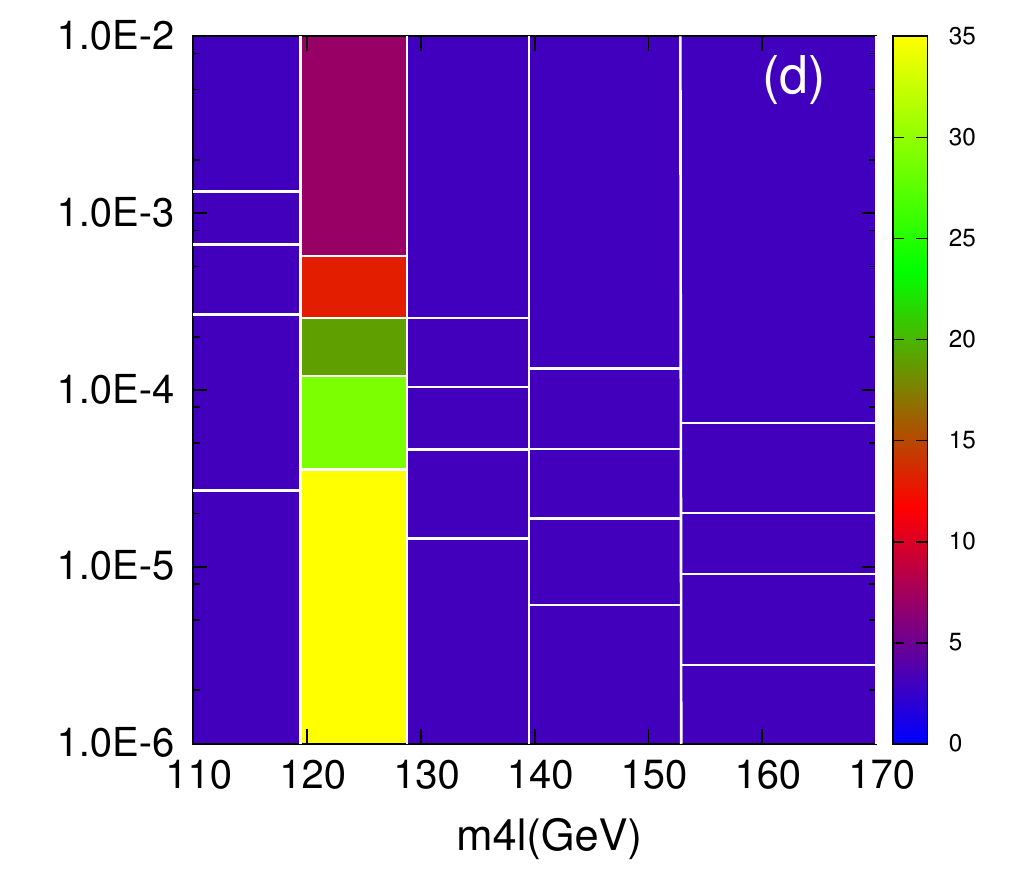}\\
\caption{\label{fig:quantile}
Quantile bins in $m_{4\ell}$ (x-axis) and $|{\mathcal M}|^2$ (y-axis) we constructed
using the background ($q\bar{q} \to 2e2\mu$) distribution.  
We then plot the number of events in each quantile bin
either from $150$ background $q\bar{q} \to 2e2\mu$ events (panels in
the top row), or $75$ $125$-GeV Higgs signal and $75$ background events
(panels in the bottom row).  The panels in the left column are for one $150$
event pseudo-experiment, while the panels in the right column
are for the average of $400$ pseudo-experiments.}
\end{figure}

\begin{figure}[t]
\includegraphics[width=0.49\columnwidth]{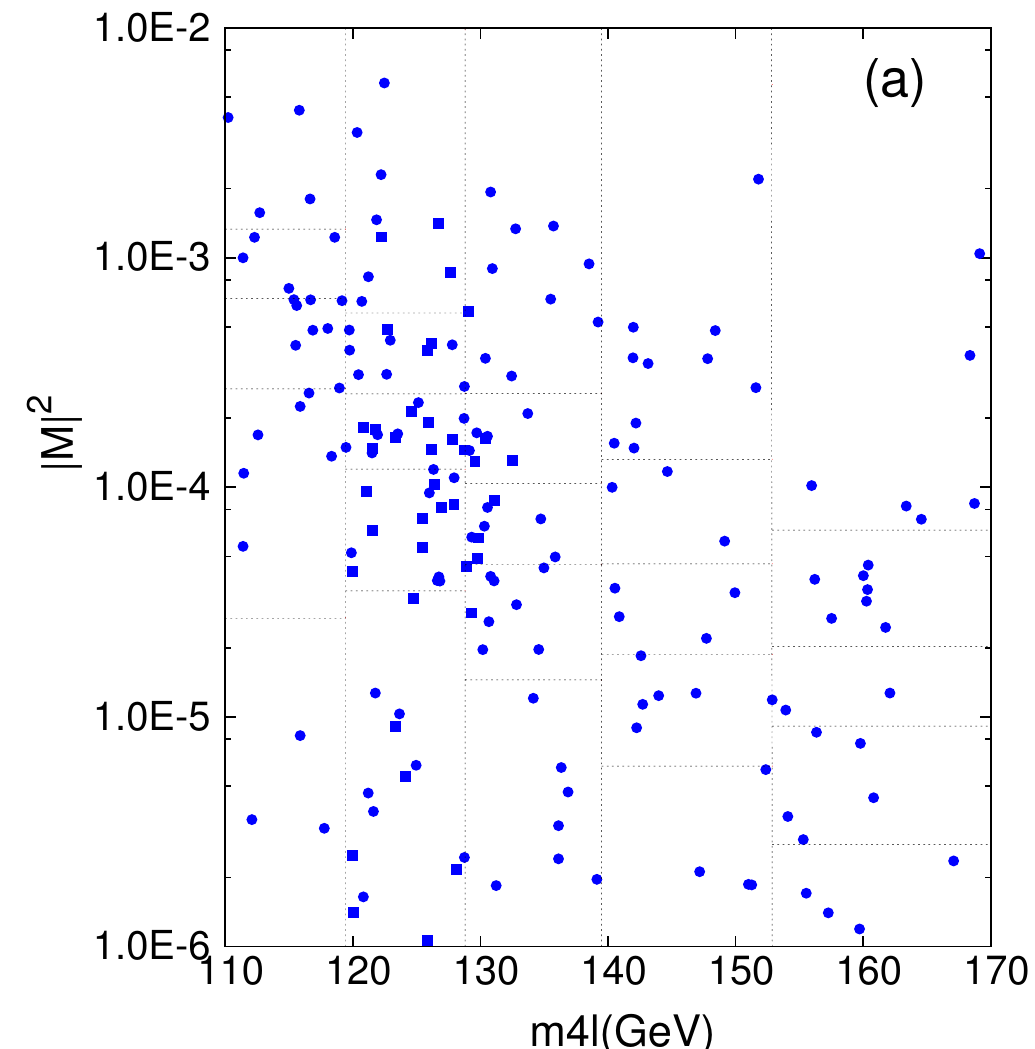} 
\includegraphics[width=0.49\columnwidth]{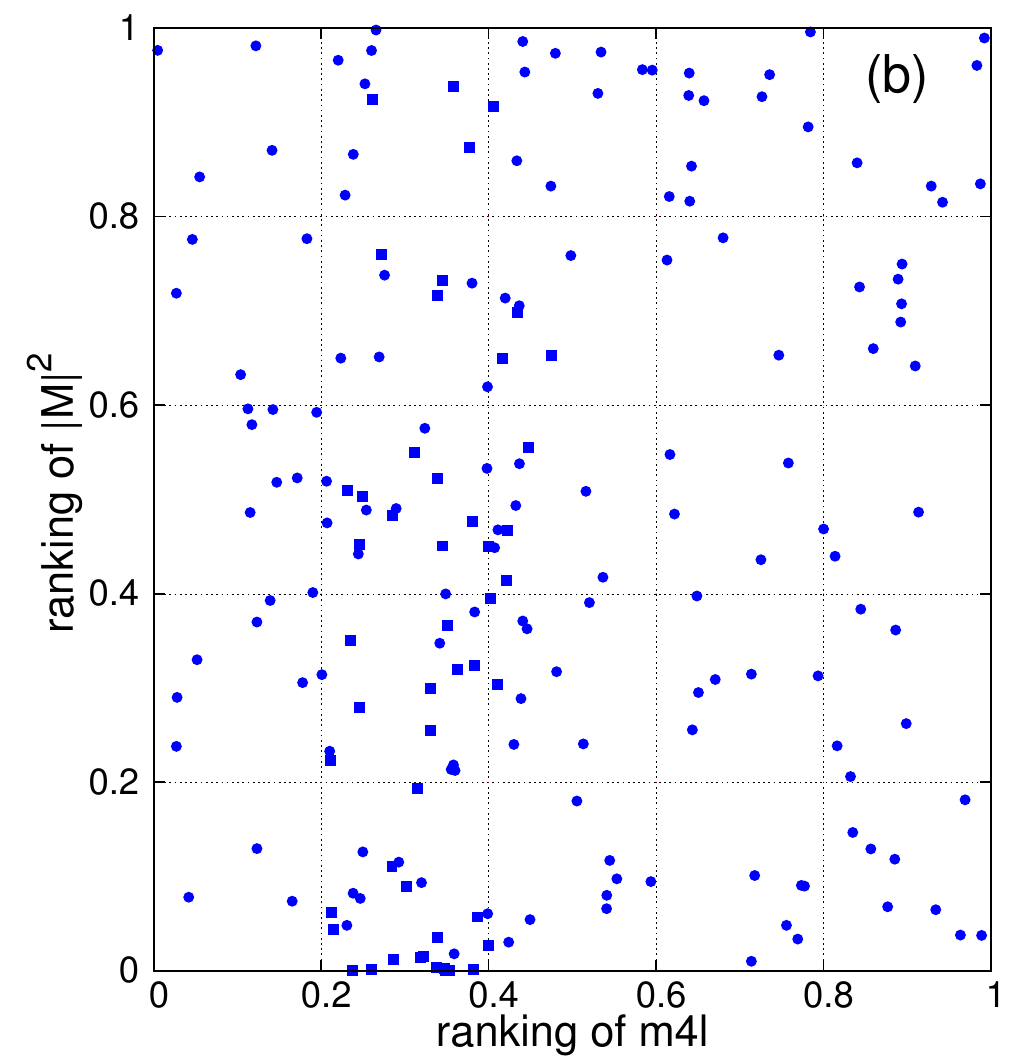} 
\caption{\label{fig:quantile-flattening}
Simulated data consisting of $50$ 125-GeV Higgs $gg \to H \to 4\ell$
events and $150$ $q\bar{q} \to 4\ell$ events.  In panel (a), the
four-lepton invariant mass and pdf-weighted background squared matrix
element have been plotted, while in panel (b) the ranking variable
corresponding to these quantities, as defined in Eq.~(\ref{eq: rank int}) is plotted.
In each case, the dotted line mesh represents the quantile bin boundaries.
}
\end{figure}

\emph{2.~~Flattening with Quantile Bins.}~~
An alternate approach is to use the method of quantile bins.\footnote{
Quantile bins have previously been employed in studies of the LHC inverse problem~\cite{AKTW}.
}
If we are only considering one variable, $\xi$, this approach consists of
finding $n+1$ values $\eta_1, \eta_2, ..., \eta_{n+1}$ such that
\begin{equation}
\int_{\eta_i}^{\eta_{i+1}} \frac{dN}{d\xi} d\xi = 1/n,
\end{equation}
i.e., the integral of the distribution is equal in each bin.
This procedure can be extended to the case where there are several
variables $\xi_i$, where again we demand that the integral of the
distribution be the same in each bin.  For example, in two dimensions,
we must choose values of $\xi_1$: $\eta_{1,1}$, $\eta_{1,2}$, ...,
$\eta_{1,n+1}$
and values of $\xi_2$: $\eta_{2,1}$, $\eta_{2,2}$, ..., $\eta_{2,n+1}$, such that
\begin{eqnarray}
\int_{\eta_{1,i}}^{\eta_{1,i+1}} \int_{\eta_{2,i}}^{\eta_{2,i+1}} \frac{d^2N}{d
  \xi_1 d \xi_2} d \xi_1 d \xi_2 = \frac{1}{n^2}.
\end{eqnarray}
This procedure allows us to consider additional kinematic variables in
addition to a likelihood-based variabe.  Examples of this are shown in
Figs.~\ref{fig:quantile} and~\ref{fig:quantile-flattening}, in which we consider the
distribution of four-lepton events at the 8 TeV LHC in terms of the
four-lepton invariant mass, $m_{4 \ell}$, and the background {\sc
  MEKD} value. 

In Fig.~\ref{fig:quantile}  we show the results of an example
experiment where we have formed quantile bins in  $m_{4\ell}$ and
$|{\mathcal M}|^2$, assuming the background hypothesis.
We then plot the number of events in each quantile bin
either from $150$ background $q\bar{q} \to 2e2\mu$ events (panels in
the top row), or $75$ $125$-GeV Higgs signal and $75$ background events
(panels in the bottom row).  The panels in the left column are for one $150$
event pseudo-experiment, while the panels in the right column
are for the average of $400$ such pseudo-experiments.
Fig.~\ref{fig:quantile-flattening} illustrates the same concept using
scatter plots.  Here the ratio of signal to background events has been
changed from 1:1 (which is realistic for $125$ GeV $H\to 4\ell$ signal
and the $q\bar{q} \to 4\ell$ background) to the much more challenging 1:3.  
Nevertheless, the presence of new signal can still be inferred from the 
anomalous clustering of points. Note that departures
from uniform density are easier to interpret in the scatter plot in
panel (b), which utilizes ranking variables.

\emph{3.~~Flattening with Respect to All the Variables.}~~
An extreme case of flattening the background distribution with 
respect to kinematic variables occurs when we consider a complete
set of kinematic variables for some process.  We can, of course, 
calculate the boundaries of these bins with Monte Carlo.
However, in the limit where we have a good analytical, or at least
numerical, understanding of the background, we can perform a 
flattening using the background distribution.

\begin{figure}[t]
\includegraphics[width=\columnwidth]{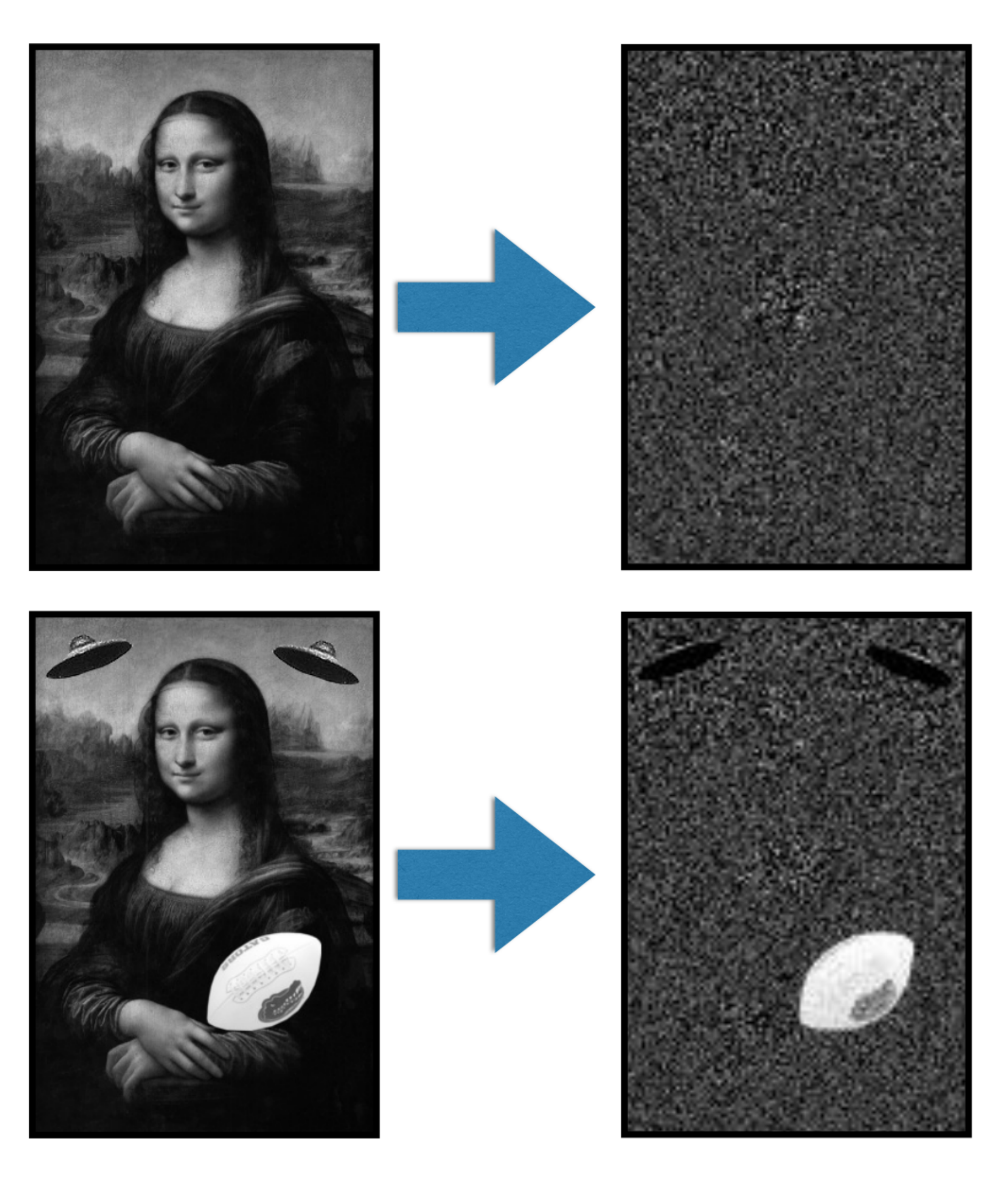}\\
\vspace{-15 pt}
\caption{\label{fig:mona}
A demonstration of filling histograms with the reciprocal of the PDF for the case of background only (top row) 
and in the presence of both signal and background (bottom row).
}
\end{figure}

Specifically, if the background (after detector simulation, etc.)
is described by the differential distribution $d^n N / d \bm{\xi}$,
then if we weight each background event by
$1/(d^n N / d \bm{\xi})$, we will end up with a distribution 
that is flat in the full $n$-dimensional space of values.  
If we weight data events according to this procedure, 
a signal will show up as deviations from flatness.  
This procedure is demonstrated in Fig.~\ref{fig:mona}.
The image in the top left represents our background PDF.  If we
generate ``events'' (i.e., pixels) according to this PDF, but
weigh the corresponding 2D histogram by the reciprocal of the PDF,
then we obtain an essentially flat distribution, shown in the top
right corner.  We now consider the bottom left image, where some
``signal'' (American football and flying saucers) have been added to
the background.  If events are generated according to this PDF, but
weighted according to the reciprocal of the background PDF, we obtain
the bottom right image, in which background features have been
flattened, but signal features remain distinct.

{\bf Conclusions.}~~We have presented methods, which utilize variables based on the
squared matrix element, to search for new physics signals at the LHC in a model
independent way.  These approaches 
allow for model-independent exclusions of the standard model in the
presence of arbitrary, unspecified, new physics.  We look forward to the utilization of such
methods in the upcoming Run 2 at the LHC.

{\bf Acknowledgements.}
J.G. and K.M. would like to thank their CMS colleagues for useful
discussions.  All of the authors would also like to thank D.~Kim and
T.~LeCompte for stimulating conversations and L. da Vinci for
assistance with Fig.~\ref{fig:mona}.
Work supported in part by U.S. Department of Energy Grant
DE-FG02-97ER41029.

\end{document}